\documentclass[useAMS,usenatbib]{mn2e}
\usepackage{times}
\usepackage{graphicx}
\usepackage{txfonts}
\usepackage{natbib}
\usepackage{tabularx}
\usepackage{nth}
\usepackage[switch]{lineno}
\usepackage{xspace}
\usepackage{bm}
\usepackage{url}

\usepackage{booktabs}

\newcommand{\vect}[1]{\ensuremath{\bm{#1}}}

\newcommand{\pvec}{\ensuremath{\vect{\theta}}\xspace}

\begin{document}
\title{Exoplanet Transmission Spectroscopy using KMOS}

\author[H. Parviainen et al.]{Hannu Parviainen,$^1$\thanks{hannu.parviainen@physics.ox.ac.uk} Suzanne Aigrain,$^1$
Niranjan Thatte,$^1$ \newauthor Joanna K. Barstow,$^1$ Thomas M. Evans,$^2$ and Neale Gibson$^3$\\
$^1$Department of Physics, University of Oxford, Keble Road, Oxford OX1 3RH, UK\\
$^2$Astrophysics Group, School of Physics, University of Exeter, Stocker Road, Exeter, EX4 4QL, UK\\
$^3$European Southern Observatory, Karl-Schwarzschild-Str. 2, D-85748 Garching bei M\"unchen, Germany}

\maketitle

\begin{abstract}
KMOS (K-Band Multi Object Spectrograph) is a novel integral field spectrograph installed in the VLT's ANTU
unit. The instrument offers an ability to observe 24 2.8\arcsec{}$\times$2.8\arcsec{} sub-fields positionable within a
7.2\arcmin{} patrol field, each sub-field producing a spectrum with a 14$\times$14-pixel spatial resolution. The main
science drivers for KMOS are the study of galaxies, star formation, and molecular clouds, but its ability to
simultaneously measure spectra of multiple stars makes KMOS an interesting instrument for exoplanet atmosphere
characterization via transmission spectroscopy. We set to test whether transmission spectroscopy is practical with KMOS,
and what are the conditions required to achieve the photometric precision needed, based on observations of a partial
transit of WASP-19b, and full transits of GJ~1214b and HD~209458b. Our analysis uses the simultaneously
observed comparison stars to reduce the effects from instrumental and atmospheric sources, and Gaussian processes to
model the residual systematics. We show that KMOS can, in theory, deliver the photometric precision required for
transmission spectroscopy. However, this is shown to require a) pre-imaging to ensure accurate centering and b) a very
stable night with optimal observing conditions (seeing $\sim$0.8\arcsec{}). Combining these two factors with the need to
observe several transits, each with a sufficient out-of-transit baseline (and with the fact that similar or better
precision can be reached with telescopes and instruments with smaller pressure,) we conclude that transmission
spectroscopy is not the optimal science case to take advantage of the abilities offered by KMOS and VLT.
\end{abstract}

\begin{keywords}
Instrumentation: spectrographs--Techniques: photometric--Techniques: spectroscopic: Planets and satellites: atmospheres
\end{keywords}

\section{Introduction}
\label{sec:introduction}
Transmission spectroscopy, the measurement of a transit depth as a function of wavelength, allows us to probe the
existence and abundance of different atmospheric species--each with their wavelength-dependent extinction features--in
planetary atmospheres~\citep{Brown2001a}. However, the variations in the transit depth are minute, and high-precision
spectroscopic time series are required in the characterization of the planetary transmission spectra. 

Systematic trends from changing telluric and instrumental conditions impair the ground-based measurements, and the
highest-quality transmission spectroscopy observations have been carried out using space-based HST until
recently~\citep{Charbonneau2002,Berta2011,Sing2011a,Sing2013,Knutson2014,Evans2013}. However, the use of multi-object
spectrographs in combination with modern data analysis methods has led to remarkable improvements in the precision that
can be achieved from the ground. Simultaneous measurements of the target star and several comparison stars allow for the
correction of common-mode systematics, in parallel to relative
photometry~\citep{Bean2010a,Rabus2013,Gibson2012a,Gibson2013,Bean2013,Murgas2014}. Further, the use of Gaussian
processes has facilitated the correction of systematics by allowing for the robust modeling of correlated
noise---including time correlation and correlations with auxiliary measurements such as seeing and humidity---in
model-independent fashion~\citep{Gibson2011a,Roberts2013,Rasmussen2006}.

The K-band Multi Object Spectrograph \citep[KMOS,][]{Sharples2013} is an integral field spectrograph installed in the
Nasmyth-focus of the VLT's ANTU unit. KMOS consists of 24 arms that can be positioned (nearly) freely inside a
7.2\arcmin{} diameter patrol field. The ability to observe multiple stars and sky fields simultaneously makes KMOS a
potentially promising instrument for transmission spectroscopy. However, the small spatial extent of IFUs and the
relative complexity of transforming the raw data to science datacubes may impair the photometric precision. The small
per-IFU FOV can especially be expected to cause problems if the seeing varies significantly during the observations.

Here we document our experiences in assessing the KMOS' applicability to transmission spectroscopy based on the 
commissioning observations of a partial transit of WASP-19b, and our observations of the full transits of GJ~1214b and 
HD~209458b.$\!$\footnote{Based on observations collected at the European Organisation for Astronomical Research in the
Southern Hemisphere, Chile. Proposals 60.A-9239(C), 60.A-9447(A), and 092.C-0812(A)}

\section{Observations}
\label{sec:observations}

\begin{table*}[htp]
  \centering
    \begin{tabularx}{\textwidth}{@{\extracolsep\fill}lllllllrrrr}
    \toprule
    Target & Starting night & Time span & Grating & N$_{\mathrm{sci}}$ & N$_{\mathrm{dark}}$ & N$_{\mathrm{flat}}$ & 
N$_{\mathrm{arc}}$ &T$_{\mathrm{exp}}$ [s] & N$_{\mathrm{ref}}$ & N$_{\mathrm{sky}}$ \\
    \midrule
    WASP-19b   & 29.03.2013 & 03:11--05:10 & HK & 64  & 5 &  4,1 & 1,1 & 10 &  6 & 6 \\
    GJ~1214b   & 29.06.2013 & 02:34--05:14 & HK & 116 & 5 & 18,3 & 6,1 & 20 & 11 & 9 \\
    HD~209458b & 28.10.2013 & 23:49--03.53 & IZ & 226 & 5 & 18,3 & 6,1 & 60 &  5 & 18 \\
    \bottomrule
  \end{tabularx}
  \caption{Observation details, where N$_{\mathrm{sci}}$ is the number of science exposures, N$_{\mathrm{dark}}$ is the 
number of dark exposures, N$_{\mathrm{flat}}$ is the number of flat field exposures (on,off), N$_{\mathrm{arc}}$ is the 
number of arc lamp exposures (on,off), T$_{\mathrm{exp}}$ is the science frame exposure time, N$_{\mathrm{ref}}$ is the 
number of reference stars, and N$_{\mathrm{sky}}$ is the number of sky fields.}
  \label{tab:observations}
\end{table*}

\subsection{Instrument}
\label{sec:observations:instrument}
KMOS \citep{Sharples2013} is installed in the Nasmyth-focus of the VLT's ANTU unit, and consists of 24 pickoff mirrors
positionable inside a 7.2\arcmin{} diameter patrol field. Each mirror covers a square 2.8\arcsec$\times$2.8\arcsec{}
field, which is fed to an image slicer that divides the field into 14$\times$14 spatial elements, each of which are
further dispersed into spectra using a grating spectrometer. The 24 arms are divided into three groups, each with its
own spectrograph. The spectra for each IFU and spatial element are arranged in the detector as one-dimensional columns,
which can be constructed into three-dimensional image cubes through steps described by \citet{Davies2013}.

The three spectrographs use substrate-removed 2kx2k 18$\mu$m-pixel Hawaii 2RG (HgCdTe) detectors (one per spectrograph)
cooled to 40K. Detector readout is done using sample-up-the-ramp mode, that is, the detector is read out continuously
without resetting it, and the counts for each pixel are computed by fitting the slope of the signal against
time.$\!$\footnote{A technical overview of KMOS can be found from
\url{https://www.eso.org/sci/facilities/paranal/instruments/kmos/inst.html}}

\subsection{WASP-19b}
\label{sec:observations:wasp-19b}

\begin{figure}
 \centering
 \includegraphics[width=\columnwidth]{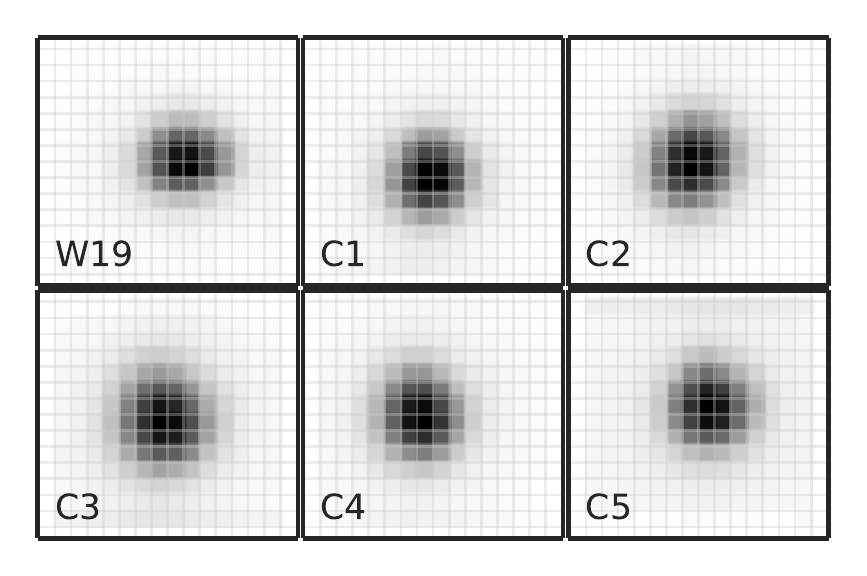}
 \caption{Log~fluxes for WASP-19b and the five best comparison stars corresponding to a single exposure where the 
datacube has been flattened in the spectral direction.}
 \label{fig:wasp_19b_ifus}
\end{figure}

A partial transit of WASP-19b \citep[H=10.6][]{Hebb2010} was observed on 29.03.2013 (03:11--05:10) as a part of the 
commissioning of the 
instrument. The observations were carried out in stare mode\footnote{Some dithered exposures were taken after the 
stare-mode observations, but these are excluded from our analysis.} using the HK grating  covering the wavelength 
ranges from 1.48~to~2.44~$\mu$m with a spectral resolution of $\sim$1800. The total exposure time was 60~s, with  
$\mathrm{DIT}=10$~s and $\mathrm{NDIT}=6$ (where DIT~=~detector on-chip integration time, NDIT~=~number of exposures
averaged over a single frame.) Six reference stars and six sky fields were observed simultaneously with 
WASP-19. All stars were sufficiently well-centered, and the PSFs are well contained inside the IFUs 
(Fig.~\ref{fig:wasp_19b_ifus}).

\begin{figure}
 \centering
 \includegraphics[width=\columnwidth]{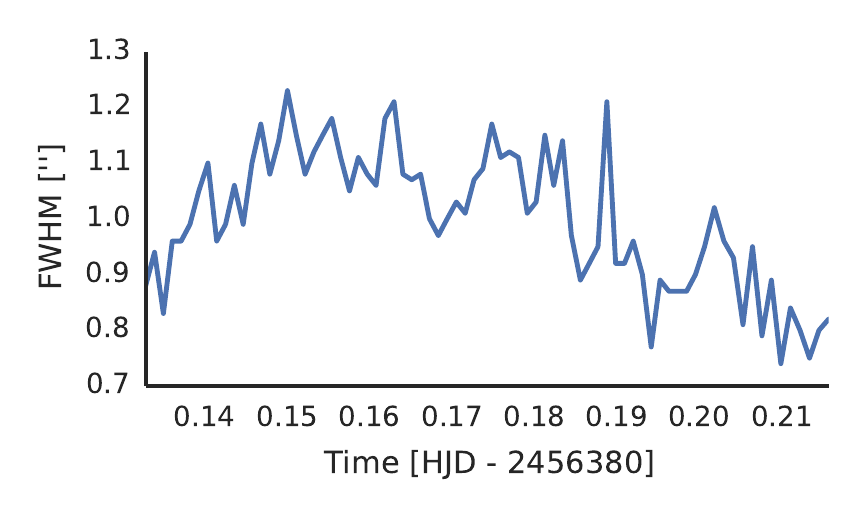}
 \caption{Seeing (as PSF FWHM) during the WASP-19b observations.}
 \label{fig:wasp_19b_seeing}
\end{figure}

The airmass was at its minimum at the beginning of the observations, increasing from 1.10 to 1.35 by the end of the 
observations. The seeing, shown in Fig.~\ref{fig:wasp_19b_seeing}, was slightly worse than the median Paranal seeing of 
(FWHM of 0.83\arcsec), varying from 0.74\arcsec to 1.23\arcsec with a median of 1.0\arcsec.

The observations cover only the last $\sim1/3$ of the transit, but include a post-transit baseline of 1.2~h. The lack 
of sufficient in-transit coverage render the observations useless in transmission spectroscopy, but they can still be 
used to estimate the precision that could have been achieved if a full transit had been observed, as detailed later in 
Sec.~\ref{sec:wasp_19b:precision_test}. 

\subsection{GJ~1214b}
\label{sec:observations:gj_1214b}

We observed a transit of GJ~1214b (H=9.1) on 29.06.2013 (2:34--5:15) using the HK grating. The total 
exposure time  was 60~s, leading to average efficiency of 72\% with DIT=20~s and NDIT=3 (average overheads were 
0.39~min/frame). The observations covered the main target GJ~1214b (arm~2), four primary reference stars (arms 
10,15,16,17), and seven secondary reference stars (arms 
3,5,6,7,11,20,22). The observations were carried out in the stare mode, and nine arms were assigned to observe the sky. 
The average FWHM during the night was 0.8\arcsec{}, close to the median Paranal seeing, with a standard deviation of 
0.1\arcsec.
One of the primary reference stars (IFU~16) was not observed due to technical issues.

\begin{figure}
 \centering
 \includegraphics[width=\columnwidth]{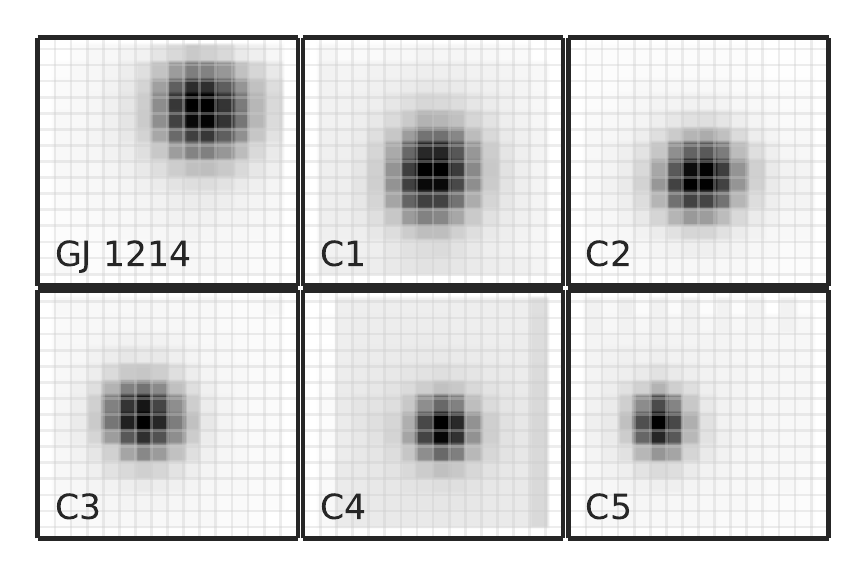}
 \caption{Log~fluxes for GJ~1214b and the five best comparison stars.}
 \label{fig:gj_1214_ifus}
\end{figure}

GJ~1214 has a high proper motion, and its position at the time of observations was calculated from the proper motion 
estimates that did not include uncertainty estimates. Unfortunately, the uncertainties were large enough to position 
the main target close to the upper-right corner of its IFU, as shown in  Fig.~\ref{fig:gj_1214_ifus}, which led to 
severe systematics due to a varying amount of flux being lost outside the IFU with varying seeing.

\subsection{HD~209458b}
\label{sec:observations:hd_209458b}

\begin{figure*}
 \centering
 \includegraphics[width=\textwidth]{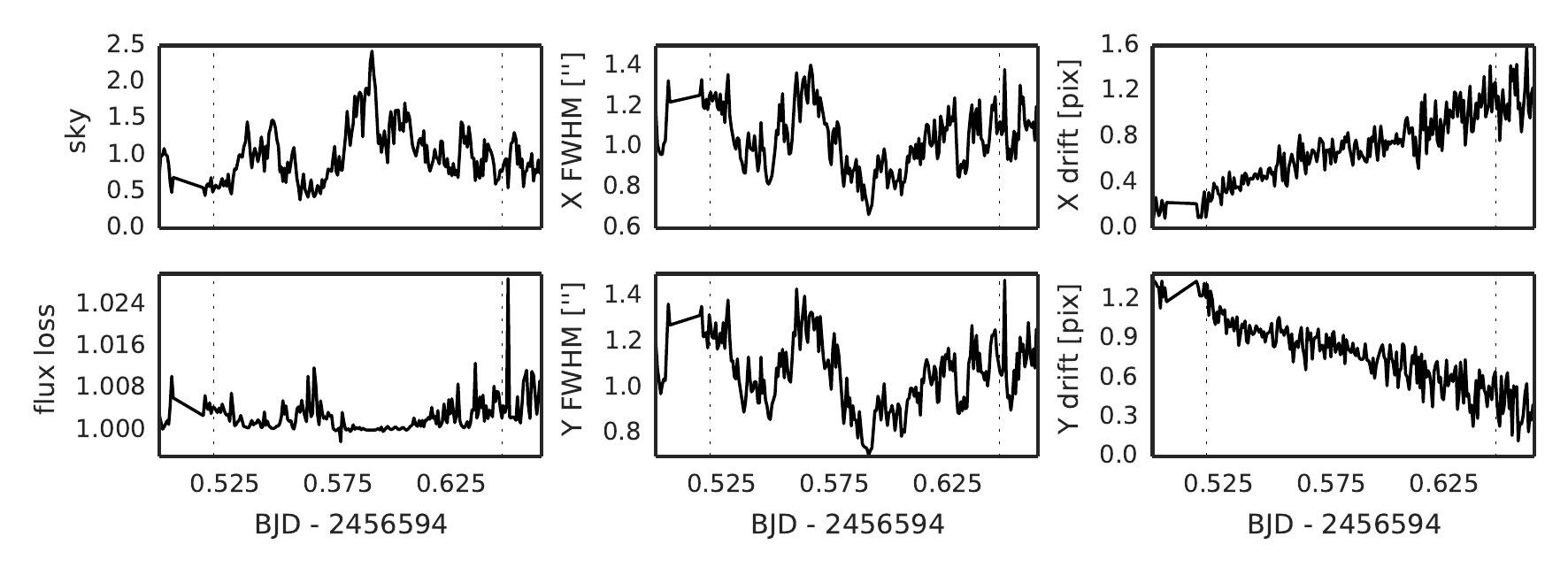}
 \caption{Auxiliary information from the instrument (FITS headers) and PSF fitting for the HD~209458b observations. The 
  focus value is from the FITS headers, while the residual sky level, x and y FWHMs and x and y drifts are from the PSF 
  fitting (averaged over the four brightest stars).}
 \label{fig:hd_209458b_aux_pars}
\end{figure*}

We observed a full transit of HD~209458b (H=6.4) on 28.10.2013 (23:49--03:53) in the stare mode using the IZ grating 
covering the wavelength range from 0.8 to 1.06~$\mu$m with a spectral resolution of $\sim$3200. The total exposure time 
was 60~s with $\mathrm{DIT}=10$~s and $\mathrm{NDIT}=6$. Five reference stars and 18 sky fields were observed 
simultaneously with HD~209458. All the reference stars were significantly fainter than the target, the three brightest 
being 3.4\%-5.6\% of the HD~209458's flux each. The airmass was at its minimum in the beginning of the observations, 
increasing from 1.4 to 2.8 by the end of the observations. The seeing was non-optimal (median FWHM for the night was 
1.06\arcsec, while the Paranal median FWHM is 0.83\arcsec), and varied from 0.75\arcsec{} to 1.8\arcsec{} during the 
night (Fig.~\ref{fig:hd_209458b_aux_pars}).

\begin{figure}
 \centering
 \includegraphics[width=\columnwidth]{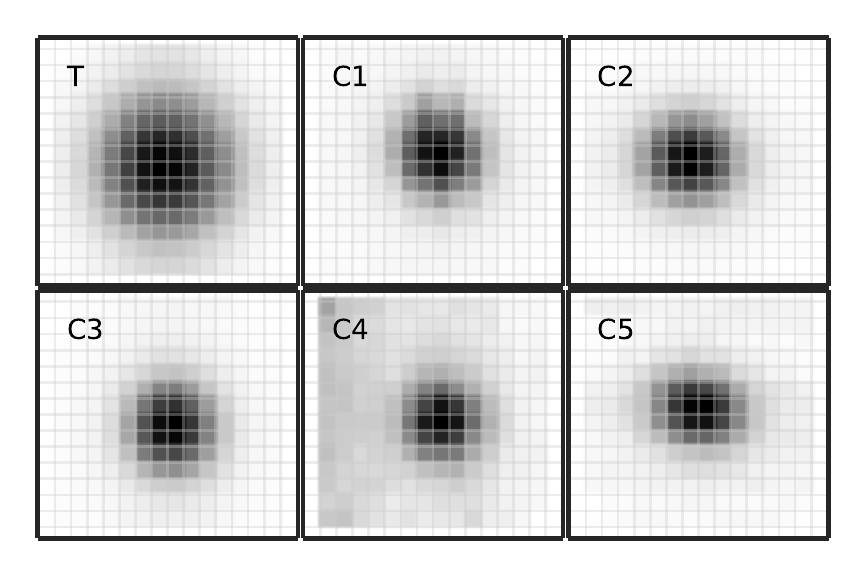}
 \caption{Log fluxes for HD~209458b (T) and five best comparison stars. The fourth comparison star (C4) was not used in 
  the analysis.}
 \label{fig:hd_209458b_target_images}
\end{figure}

All the stars were well-centered on their IFUs (Fig.~\ref{fig:hd_209458b_target_images}), and we did not run into 
similar problems as with our observations of GJ~1214b. However, only a minimal pre-transit baseline was observed due to 
technical difficulties. Also, the post-transit baseline was shorter than optimal due to observing limitations. This 
lack of adequate out-of-transit baseline can be expected to have a strong impact on the precision of the transit depth 
estimates.

\section{Data reduction}
\label{sec:reduction}
\subsection{Overview}
\label{sec:reduction:overview}

The official ESO KMOS-1.2.4 pipeline was used for the basic reductions \citep[detailed in][]{Davies2013}, following the
steps outlined in the "SPARK INstructional Guide for KMOS data" version 0.7 \citep{Fairley2012}. The pipeline uses the
calibration frames with the raw science frames to construct separate datacubes for each observed IFU, which were used in
the following analysis. The basic calibrations consist of creation of master dark and flat frames, identification of hot
and cold pixels, dark frame subtraction, and division by the master flat. The flat and arc sets consists of
N$_\mathrm{on}$ exposures with the calibration lamp on, and N$_\mathrm{off}$ exposures with the calibration lamp off,
for improved dark current removal. Hot pixels are identified from the dark frames, and cold pixels from the flat frames,
and both are excluded from the rest of the analysis. A wavelength solution is computed using the arc frames, which is
then used in the datacube creation. Cosmic ray detection and correction was not included into the basic reductions, and
neither was the sky removal, which was carried out using proprietary methods. The number of calibration frames used in
the analyses are given in Table~\ref{tab:observations}.

We experimented with four approaches to photometry and two approaches to sky estimation. Since the observations were
carried out in the stare mode, each target star was assigned with a nearby sky IFU (typically 0.4\arcmin{}--0.7\arcmin{}
from the object IFU) for sky subtraction. Thus, the photometry was carried out using a target and a sky cube for each
target and exposure. The basic steps were similar for most approach combinations, PSF fitting excluded: 
\begin{enumerate}
 \item Calculate the 1D target spectrum by collapsing the datacube in the spatial dimensions (sum the pixel values) 
       after possible masking.
 \item Calculate the 1D sky spectrum from either a simultaneously measured sky-cube or from target-cube pixels where 
       the target flux does not significantly contribute to the total flux.
 \item Subtract sky spectrum from the target spectrum.
 \item Collapse the 1D sky-subtracted target spectrum after applying a bandpass filter to obtain a single photometric 
data point.
\end{enumerate}

\subsection{Sky estimation}
\label{sec:reduction:sky_estimation}
We tried two approaches to sky estimation:
\begin{enumerate}
 \item \textbf{Using the sky IFUs.}  A 1D sky spectrum was constructed by collapsing the target-specific sky-datacube 
in spatial dimensions with median operator (that is, calculating the median over the spatial pixels for each wavelength 
  element). This spectrum was then subtracted from the 1D target spectrum.
 \item \textbf{Using the target IFUs.} A 1D sky spectrum was constructed from the target-cube pixels where the target 
    star did not significantly contribute to the observed flux (based on 2D S/N maps created for optimal-mask 
    photometry). 
\end{enumerate}
The first approach yielded the best sky removal for the bright stars, but was not optimal for the faint stars. The 
second approach yielded significantly better sky subtraction for the faint stars, but was not applicable for the bright 
stars for which all the pixels had a significant contribution from the stellar flux.

\subsection{Photometry}
\label{sec:reduction:photometry}

We experimented with four different approaches on photometry. The major difference between the three first was the 
shape of the spatial mask applied to each datacube. The masks were  constructed as 14$\times$14 real arrays with values 
ranging from 0 to 1, and the masking was done by multiplying the datacube with the mask (along the wavelength axis). 
The approaches were:

\begin{enumerate}
 \item \textbf{No masking,} where the datacubes were collapsed first in the spatial dimensions without any masking. 
    Next, the resulting 1D spectrum was multiplied with a smooth-edged window function (to select the desired 
    passband), and collapsed into a photometric point.
 \item \textbf{Soft aperture photometry}, where we used a soft-edged circular aperture mask centered on the star. The 
    shape of the mask edge followed a smoothstep function
 \begin{equation}
  \alpha = \max\left(0, \min\left(3r_N^2 - 2r_N^3, 1\right)\right),
 \end{equation}
    where $\alpha$ is the mask opacity, $r_N$ is the distance from the aperture center divided by the aperture radius, 
    and the $\min$ and $\max$ clamp the opacity to values between 0 and 1. The photometry was carried out as in the 
    first approach, but the datacube pixels were first multiplied with the aperture mask (in the x,y-space).
 
    \item \textbf{Optimal mask photometry}, where photometry masks were created for each target IFU based on the 
    (x,y)-pixel values averaged over the wavelength axis and all individual exposures. We created masks that 
    maximize the S/N ratio, but also experimented with somewhat larger masks. One mask per target was used for all 
exposures to 
    minimize the scatter from varying mask shape.
 
 \item \textbf{PSF fitting}, where an analytic PSF model was fitted to the observed data. Here we first collapsed the 
  datacubes in the wavelength axis (after applying the passband mask), and fitted the PSF to the resulting 2D image. We 
  tested two-dimensional (different FWHMs in x and y) Gaussian, Moffat, and two-component Gaussian PSF models to study 
  whether one of the models would be superior to the others. 
\end{enumerate}
The first approach (no masking) yielded the lowest scatter in the resulting light curves for the bright stars, and 
masking improved the outcome for the fainter stars. The photometry calculated from PSF fitting had significantly higher 
scatter than any of the other approaches (for all the used PSF models). However, the PSF parameter estimates (centers, 
FWHMs, residual sky level) proved to be useful in the following Gaussian Process-based detrending, since they allow us 
to model the flux loss, which was the main cause of the strongest systematics in our data.

\section{Analysis}
\label{sec:analysis}
\subsection{Overview}
\label{sec:analysis:overview}

We detail the dataset-specific analyses for WASP-19b, GJ~1214b, and HD~209458b in Sects.~\ref{sec:wasp_19b}, 
\ref{sec:gj1214b}, and \ref{sec:hd_209458}, respectively and review here the common parts.

The transmission spectroscopy is based on Bayesian parameter estimation, where we obtain a sample from the joint 
posterior distribution using Markov chain Monte Carlo (MCMC). The parameter estimates correspond to the marginal 
posterior medians, and their uncertainties to the 68\% central posterior intervals. Most of the parameters are strongly 
constrained by informative priors based on previous studies in all the three cases covered, with the passband-specific 
planet-star radius ratios relative to the average broadband radius ratio (or planet-star area-ratios) being the most 
important unconstrained parameter of interest. 

The noise in photometric time series is rarely white \citep{Pont2006}, and we use Gaussian processes 
\citep[GPs,][]{Gibson2011a,Roberts2013,Rasmussen2006} to model the correlated noise (together with a case-by-case 
selected set of simultaneously observed auxiliary parameters used as GP input parameters).

\subsection{Priors}
\label{sec:analysis:priors}
All of the planets in our study are well characterized by extensive amounts of previous studies using transit, 
occultation, and radial velocity observations. This allows us to set tight informative priors on the planet's orbital 
parameters and average radius ratios over wide passbands (indeed, given the quality of our data, we must set tight 
priors on the orbital parameters). We list the used priors in the analysis section for each planet.

\subsection{Model}
\label{sec:analysis:model}
We model the broadband flux simultaneously with the narrow-band fluxes covering the wide passband, with the aim of 
estimating the narrow-band radius ratios (or area ratios) relative to the average broadband radius ratio. For WASP-19b 
and GJ~1214b we carry out the modeling for H and K bands separately, dividing the wide passbands into six sub-bands, 
and for HD~209458b we carry out the modeling for $i$ and $z$ bands (again with six narrow passbands).

The broadband light curve and the narrow-band light curves show OOT (out-of-transit) scatter on the same scale for all
passbands. Thus, we can assume that the noise is dominated by a wavelength-independent systematic component and not by a
photon noise. We model the systematic, wavelength-independent, component using the residuals from the simultaneous
broadband modelling (similar to the often used divide-by-white approach), but in a way that also marginalizes over the
uncertainties in the broadband modelling. The narrow-band light curves are divided by the broadband light curve scaled
by a scaling factor that is a free parameter in the model (removing the truly constant systematics component) in order
to obtain a time series of relative narrow-wide passband fluxes. From here, we have the option of using Gaussian
processes (GPs) to model the residual systematics (that is, the systematics component that is not constant over the wide
passband), using a set of simultaneously observed auxiliary parameters (or information derived from other processes,
such as PSF fitting) as inputs to the GP, or, if no strong correlations between any of the inputs and the residual
relative fluxes are found, we can assume the noise to be white and use the standard likelihood equation for normally
distributed errors, described below.

We further assume that the stellar limb darkening is constant across the wide passband (which is a simplification), and 
use a quadratic limb darkening model with informative priors based on \citet{Claret2013} limb darkening tabulations.

\subsection{Posterior and likelihood}
\label{sec:analysis:posterior_and_likelihood}
The unnormalized log posterior for our model is
\begin{equation}
 \ln P(\pvec|D) = \ln P(\pvec) + \ln P(D|\pvec),
\end{equation}
where $\ln P(\pvec)$ is the log prior for the parameter vector $\pvec$. The second term, log likelihood for our data, is
\begin{equation}
 \ln P(F|\pvec) = \ln P(F_\mathrm{W}|\pvec) + \sum_{i=1}^{N_\mathrm{pb}} \ln P(F_i|\pvec), 
\label{eq:wasp_19b_logl}
\end{equation}
where the first term is the likelihood for the broadband data, and the second term is a sum over the narrow-band
flux ratio log likelihoods. The likelihood for the broadband data follows either from the GP, or assumes normally 
distributed uncorrelated noise, which leads to the usual log likelihood equation 
\begin{equation}
 \ln P(W|\pvec) = -N \ln \sigma_\mathrm{W} - \frac{\ln 2\pi}{2} - \sum_{j=1}^N \frac{\left(W_\mathrm{j} 
-M_\mathrm{W,j}\right )^2}{2\sigma_\mathrm{W}^2},
\end{equation}
where $W$ is the observed normalized broadband flux, $M_\mathrm{W}$ is the modeled broadband flux, $N$ is the number 
of exposures, and $\sigma_\mathrm{H}$ the average broadband datapoint uncertainty. The likelihoods for the narrow-wide 
flux ratios also assume normally distributed white noise, yielding a log likelihood 
\begin{equation}
 \ln P(F_\mathrm{i}|\pvec) = -N \ln \sigma_\mathrm{r} - \frac{\ln 2\pi}{2} - \sum_{j=1}^N 
\frac{\left(\frac{\alpha F_\mathrm{i,j}}{1+\beta\left(W_\mathrm{j}-1\right)} - 
\frac{M_\mathrm{i,j}}{1+\beta\left(M_\mathrm{W,j}-1\right)}\right )^2}{2\sigma_\mathrm{r}^2},
\label{eq:logl_relative_flux}
\end{equation}
where $F_\mathrm{i}$ is the observed normalized flux for a narrow passband $i$, $\sigma_\mathrm{r}$ is the flux ratio 
scatter, $\alpha$ is the constant baseline level for the flux ratio, and $\beta$ is a scaling factor applied to both 
observed and modeled broadband flux. 

The approach is similar to the often-used method of first fitting the wide passband and subtracting the residuals from 
the narrow-band light curves, but slightly more robust, since we are marginalizing over the baseline and scale 
parameters $\alpha$ and $\beta$, and modeling the relative flux explicitly.

\subsection{Numerical methods}
\label{sec:analysis:numerical_methods}
The transit light curves were modeled using PyTransit,$\!$\footnote{Available from \url{github.com/hpparvi/PyTransit}.} 
a Python package implementing Mandel-Agol and Gim\'enez transit models optimized for transmission spectroscopy. The 
MCMC sampling was carried out with emcee \citep{Foreman-Mackey2012} a Python implementation of the affine 
invariant MCMC sampler by \cite{Goodman2010}. The sampler was initialized using a population of parameter vectors 
clumped around the local posterior maximum using PyDE,$\!$\footnote{Available from \url{github.com/hpparvi/PyDE}.} a 
Python implementation of the Differential Evolution global optimization algorithm \citep{Storn1997}

The analysis also uses the large set of tools build around SciPy and NumPy \citep{VanderWalt2011}: IPython 
\citep{Perez2007}, Pandas \citep{Mckinney2010}, matplotlib \citep{Hunter2007}, 
seaborn,$\!$\footnote{\url{stanford.edu/~mwaskom/software/seaborn}} PyFITS,$\!$\footnote{PyFITS is a product of the
Space 
Telescope Science Institute, which is operated by AURA for NASA} and F2PY \citep{Peterson2009}. The Gaussian Processes 
were computed using George.$\!$\footnote{\url{Available from \url{dan.iel.fm/george}}} \citep{Ambikasaran2014}

\section{WASP-19b: partial transit observed during the KMOS commissioning}
\label{sec:wasp_19b}
\subsection{Overview}
\label{sec:wasp_19b:overview}
The partial WASP-19b transit observed during the KMOS commissioning cannot be considered for a serious transmission 
spectrum analysis due to a lack of in-transit coverage. Nevertheless, since the observation conditions were stable, it 
can be used to assess the precision of the transit depth estimates that can be achieved with the KMOS.

The observations covered only the last third of the transit, and alone cannot constrain any of the relevant properties
(radius ratio, impact parameter, transit duration, etc.) However, since WASP-19b has been extensively studied in 
transit \citep{Hebb2010,Bean2013,Mancini2013,Mandell2013,Huitson2013,Tregloan-Reed2012}, occultation 
\citep{Gibson2010,Mancini2013,Anderson2013a}, and RVs \citep{Hellier2011a}, the planet's geometric and orbital 
parameters can be strongly constrained using informative priors, listed in Table~\ref{table:wasp_19b_priors}.

\subsection{Results from the partial transit analysis}
\label{sec:wasp_19b:true_analysis}
\begin{figure}
 \centering
 \includegraphics[width=\columnwidth]{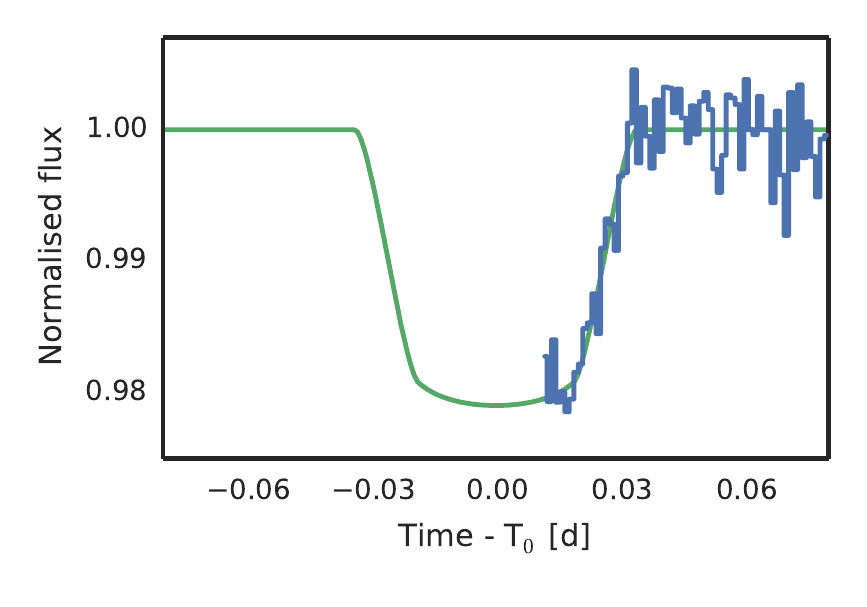}
 \caption{WASP-19b H-band transit light curve with the fitted model.}
 \label{fig:wasp_19b_lc_h}
\end{figure}

\begin{figure*}[t!]
 \centering
 \includegraphics[width=\textwidth]{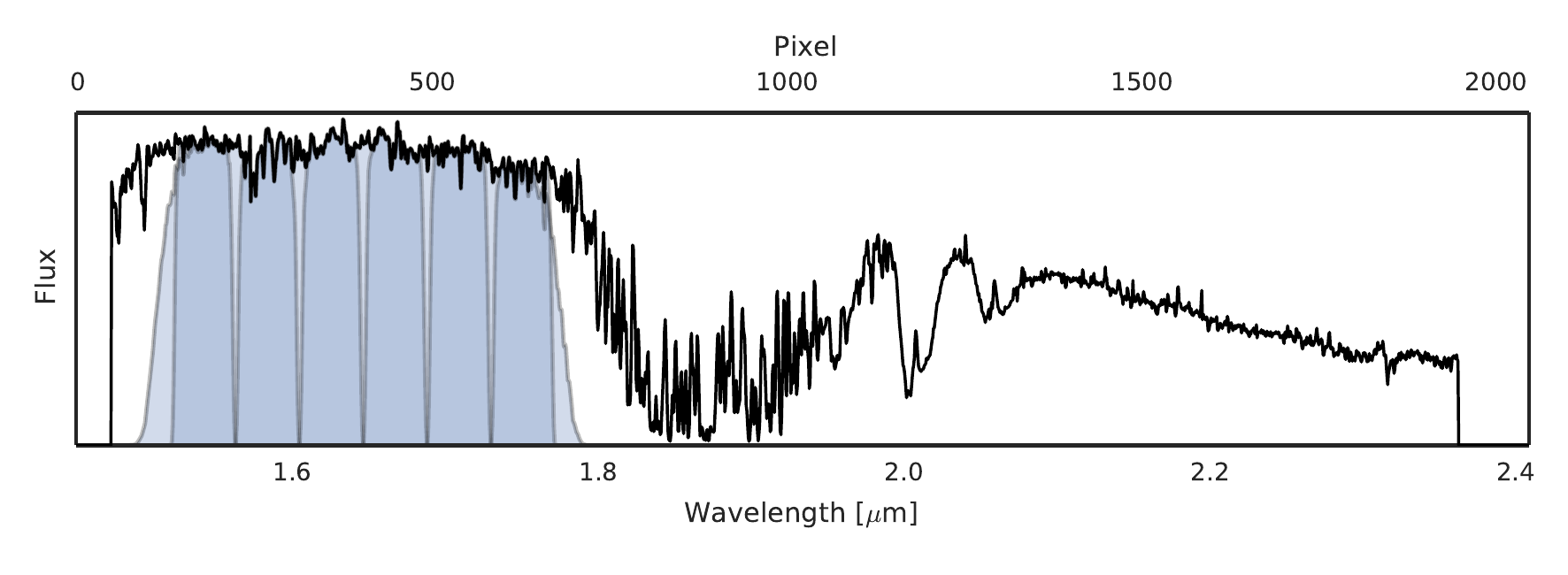}
 \caption{WASP-19 HK spectrum with the wide H-band marked as light blue and the narrow bands marked as darker blue.}
 \label{fig:wasp_19b_spectrum_h}
\end{figure*}

We carry out the WASP-19b analysis using a broadband light curve covering most of the H-band (roughly from 1.5~$\mu$m to
1.8~$\mu$m, $\sim$700~pixels), shown in Fig.~\ref{fig:wasp_19b_lc_h}, and six narrow-band light curves covering 43~nm
($\sim$100~pixels) each, as shown in Fig.~\ref{fig:wasp_19b_spectrum_h}. We detail only the H-band analysis, since the
K-band results are qualitatively similar. 

The final light curves were created by dividing the \mbox{WASP-19} light curve with the sum of the six reference star
light curves (this was found to yield the lowest final out-of-transit (OOT) ptp-scatter.) The OOT scatter of the final
normalised broadband light curve was 2.3$\times 10^{-3}$, while the theoretical shot noise was 2.4$\times 10^{-4}$. The
systematics were tested not to be correlated with any of the simultaneously measured auxiliary parameters, and we
decided not to use Gaussian processes with this dataset.

\begin{table}
\begin{tabularx}{\columnwidth}{@{\extracolsep\fill}llcl}
\toprule
Parameter & Unit & Source & Prior\\
\midrule
Orbital period & days & a & N(0.7888391, 1.1$\times 10^{-7}$)\\
Average area ratio & $A_\star$ & a & N(0.021, 0.001)\\
Impact parameter & $R_\star$ & a & N(0.681, 0.008) \\ 
Semi-major axis & $R_\star$ & b & N(3.552, 0.093) \\
Limb darkening a & & c & N(0.138, 0.010) \\
Limb darkening b & & c & N(0.252, 0.020) \\
\\
\multicolumn{4}{l}{\textit{Parameters per narrow passband with uninformative priors}}\\
Relative radius ratio & & & \\
Flux ratio baseline   & & & \\
Flux ratio scatter    & & & \\
\bottomrule
 \end{tabularx}
\caption{Informative priors used in the analysis of WASP-19b partial transit. Sources: a)~\citet{Bean2013}, 
b)~\citet{Hellier2011a}, c)~\citet{Claret2013}.}
\label{table:wasp_19b_priors}
\end{table}

\begin{figure}
 \centering
 \includegraphics[width=\columnwidth]{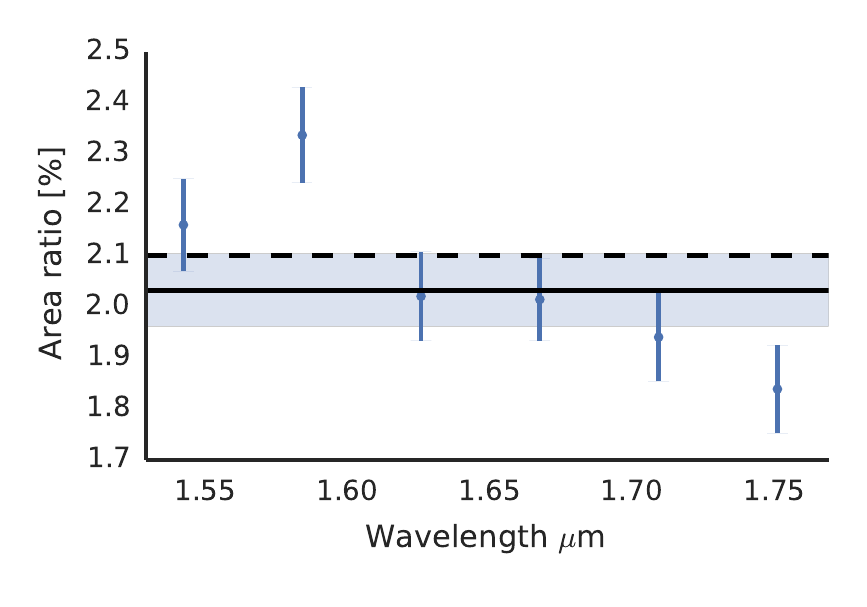}
 \caption{WASP-19b transmission spectrum estimated from the partial transit. The dots show the area ratio posterior 
medians, the errorbars correspond to the 68\% posterior central intervals, the continuous black line shows the
posterior 
median for the wide H-band area ratio, the shaded region its 68\% posterior central interval, and the slashed black
line 
the prior mean.}
 \label{fig:wasp_19b_h_transmission_spectrum}
\end{figure}

We show our results for the six narrow bands in H in Fig.~\ref{fig:wasp_19b_spectrum_h}. The 68\% posterior central 
interval widths are around 0.1\%, which is promising considering observing full transits with sufficient pre-ingress 
and post-egress baselines. However, the variation in the estimates is greater than what expected from theory (or 
measured by others, such as \citealt{Bean2013}), and thus we must consider uncertainties significantly underestimated.

\subsection{Precision test with a mock dataset}
\label{sec:wasp_19b:precision_test}

\begin{figure}
 \centering
 \includegraphics[width=\columnwidth]{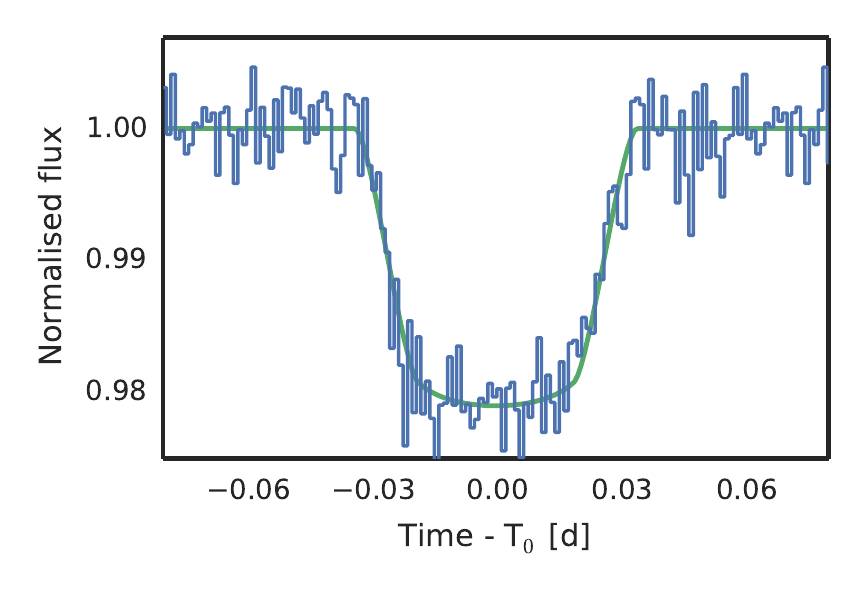}
 \caption{Mock WASP-19b H-band transit light curve with the fitted model.}
 \label{fig:wasp_19b_mock_lc_h}
\end{figure}

While it was highly unlikely from the beginning that the analysis above would yield useful results, we can
still use the observations to test the precision we could get if we would have observed a full transit with proper
baseline. We use the model fitted to the wide H-band, the wide H-band residuals, and the narrow band residuals to create
a mock dataset with similar time resolution as with the original observations. We model the broadband data by summing
the broadband model and the broadband residuals (tiled periodically to cover the whole time-span), and show the
resulting light curve in Fig.~\ref{fig:wasp_19b_mock_lc_h}. Next, we create the narrow-band light curves by summing a
narrow-band model and the narrow band residuals for each band. We set four of the narrow passbands to have exactly the
same transit depth as the wide band (passbands 1,2,3, and 6), while the fourth narrow band is set to have the area ratio
of 1.95\% and the fifth an area ratio of 2.11\%. 

\begin{figure}
 \centering
 \includegraphics[width=\columnwidth]{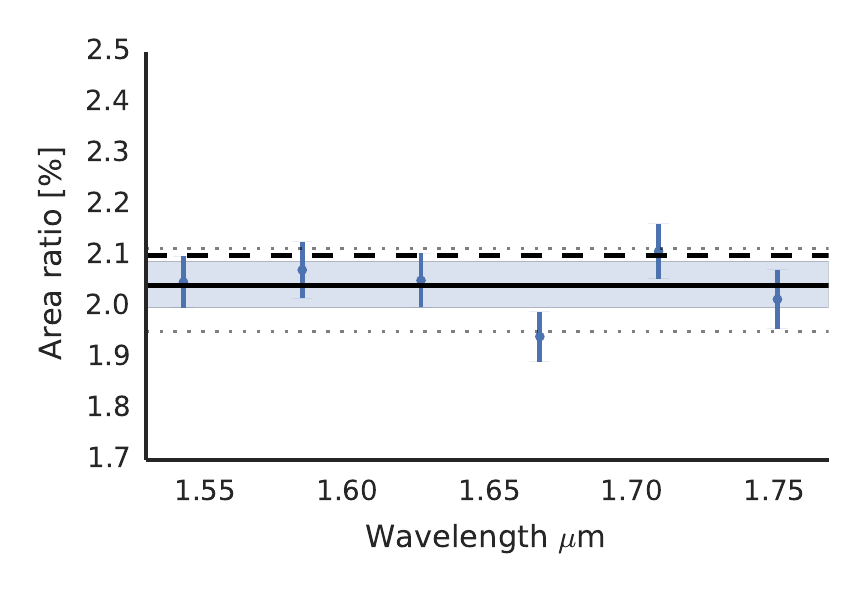}
 \caption{WASP-19b transmission spectrum from the mock dataset. The dots show the area ratio posterior medians, the 
  errorbars correspond to the 68\% posterior central intervals, the continuous black line shows the wide H-band area 
  ratio posterior median, the shaded region its 68\% posterior central interval, and the slashed black line the prior 
  mean. The dotted lines show the area ratio values set to fourth and fifth passbands, while all other passbands have 
  the same area ratio as the wide passband.}
 \label{fig:wasp_19b_h_mock_transmission_spectrum}
\end{figure}

We show the transmission spectrum obtained using the mock data in Fig.~\ref{fig:wasp_19b_h_mock_transmission_spectrum}. 
The area ratio estimates match well the input data, and the average 68\% area ratio posterior interval width 
is~$\sim$0.05\%, half of its value for the observed partial dataset.

Based on this test, we can be cautiously optimistic about using KMOS in transmission spectroscopy, given that we get 
sufficient pre-ingress and post-egress baseline. However, a note must be taken that the test above underestimates the 
effects from white noise, since repeating the residuals also repeats the white noise component.

\section{GJ 1214b}
\label{sec:gj1214b}
\subsection{Overview}

\begin{figure}
 \centering
 \includegraphics[width=\columnwidth]{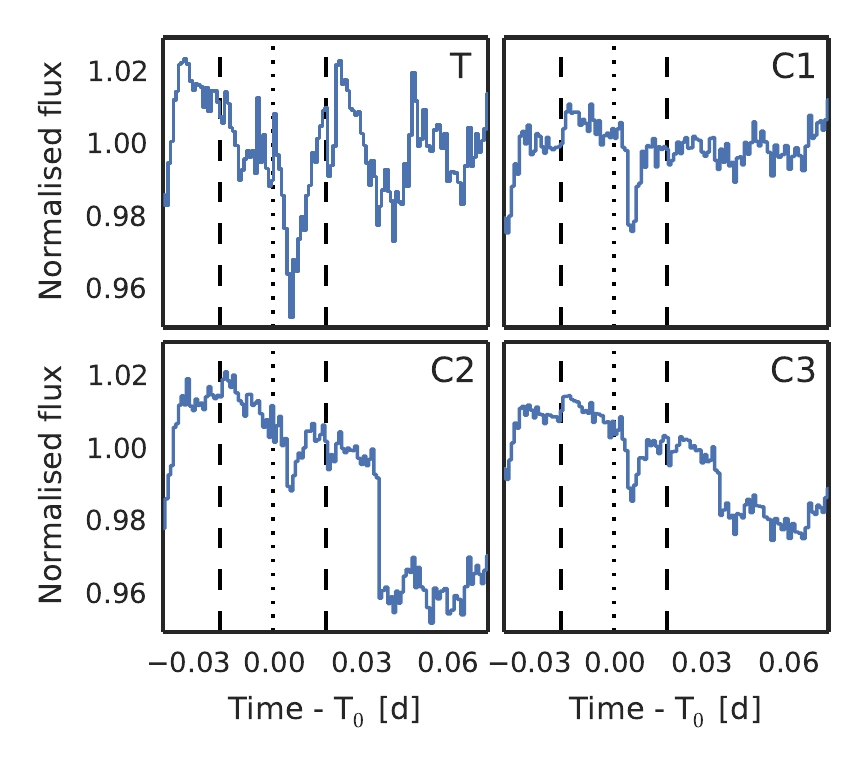}
 \caption{H-band light curves for GJ~1214 (T) and the three best comparison stars (C1-C3). The expected 
transit start, center, and end times are marked as vertical dotted lines. Both C2 and C3 feature a sharp jump during 
the 
second half of the observations.}
 \label{fig:gj_1214_h_raw_lcs}
\end{figure}

\begin{figure}
 \centering
 \includegraphics[width=\columnwidth]{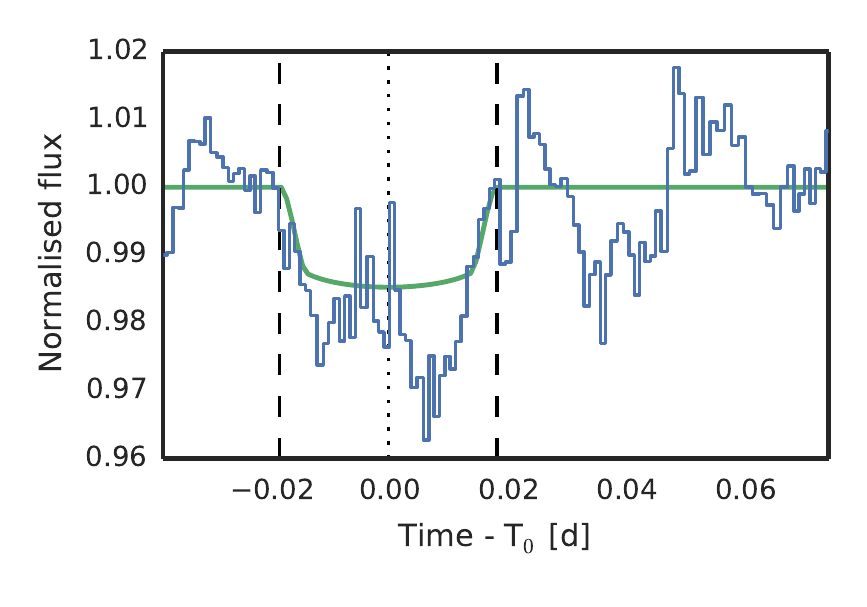}
 \caption{GJ~1214b H-band transit light curve with a model based on parameter estimates from \citet{Kreidberg2013}. The 
dotted line shows the predicted transit center, and the dashed lines the beginning of ingress and the end of egress.}
 \label{fig:gj_1214_lc_h}
\end{figure}

The GJ~1214b observations suffered from inaccurate centering. The host star has a high proper motion, and it ended up 
being positioned close to the a corner of the IFU, as shown in Fig.~\ref{fig:gj_1214_ifus}.
The poor centering leads to strong seeing-related systematics due to varying amounts of flux being lost outside the 
IFU, as seen in Fig.~\ref{fig:gj_1214_h_raw_lcs}. Dividing the target star light curve with the sum of the three best 
comparison star light curves mitigate the systematics (Fig.~\ref{fig:gj_1214_lc_h}), but not sufficiently. One reason 
for this is that the flux loss is non-linear and depends on the location of the PSF center in the IFU (the 
elliptically symmetric PSF is bounded by a rectangular area). The systematics from flux loss can be mitigated by 
modeling the flux loss for each star based on PSF fitting, as carried out for the HD~209458 observations discussed 
later in Sect.~\ref{sec:hd_209458}, but the approach was not sufficient for this dataset.

Further, the light curves of comparison stars C1 and C2 in Fig.~\ref{fig:gj_1214_h_raw_lcs} show a sudden jump during 
the after-transit baseline observations. This jump relates to a sudden shift in the spectrum (in wavelength dimension) 
in some of the IFUs. The reason for the shift is unknown, and likely arises during the generation of datacubes from 
the raw data. The shift occurs at the same position for all IFUs where it is noticeable, and happens  in detectorss 1
and 2. The photometric signal arises from the subtraction of a target IFU spectrum featuring a wavelength shift with a
sky IFU spectrum without the shift.

\subsection{Analysis}
\label{sec:gj_1214b:analysis}

\begin{figure*}[t!]
 \centering
 \includegraphics[width=\textwidth]{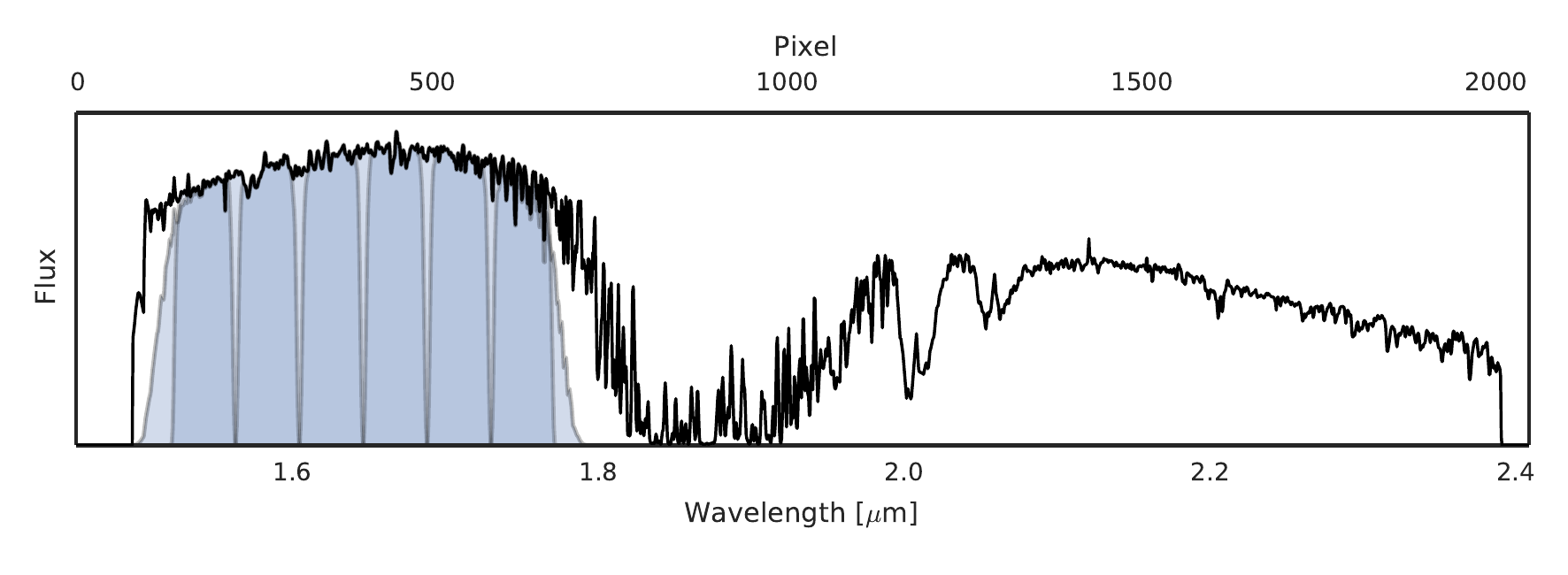}
 \caption{GJ~1214 HK spectrum with the wide H-band marked as light blue and the narrow bands marked as darker blue.}
 \label{fig:gj_1214_spectrum_h}
\end{figure*}

The strong systematics render the dataset useless in basic characterization, and more so in transmission spectroscopy. 
The transit is barely recognizable from the broadband relative photometry light curve (Fig.~\ref{fig:gj_1214_lc_h}), 
and the systematics are not constant in wavelength. The measured broadband light curve rms scatter is
$\sim 8\times10^{-3}$, while the theoretical shot noise level is $\sim 1.3\times10^{-4}$.

We nevertheless investigate whether Gaussian processes \citep{Gibson2011a,Roberts2013,Rasmussen2006} can be used to 
model the systematics based on a set of auxiliary parameters (pressure, seeing, sky level), and the modeled flux loss 
based on PSF fitting. We condition a GP with an squared exponential kernel to the out-of-transit points (with a 
separate input-scale for each parameter) after first optimizing the GP likelihood as a function of kernel 
hyperparameters. The GP can explain some of the variability, but not to a sufficient level for a meaningful analysis.

We further carry out an analysis similar to the WASP-19b transmission spectroscopy analysis. We set very narrow 
priors on the orbital parameter, limb darkening, and average radius ratio, based on the results by 
\citet{Kreidberg2013}, and set to estimate the narrow-band transit depths over the H-band based on 
Eq.~\ref{eq:logl_relative_flux}. 

\begin{figure}
 \centering
 \includegraphics[width=\columnwidth]{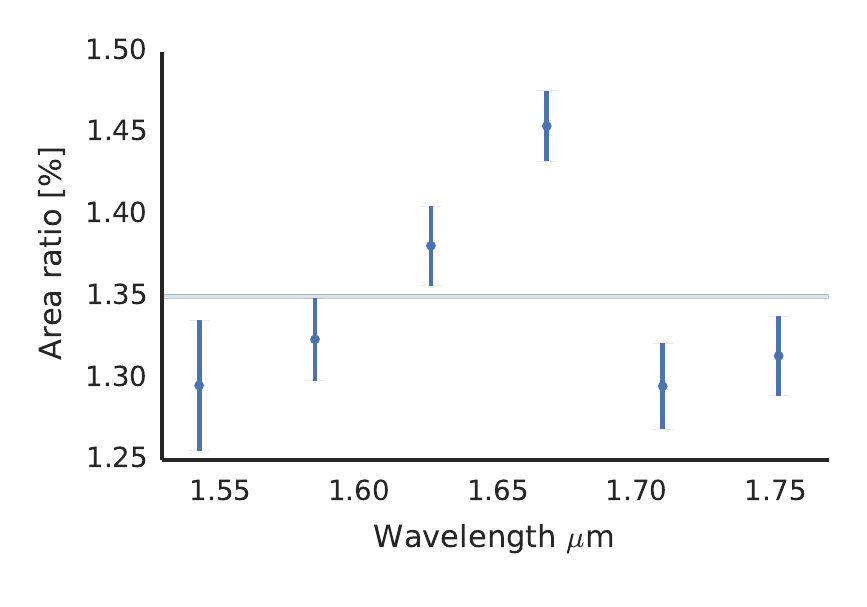}
 \caption{GJ~1214b H-band transmission spectrum estimated from the data. The light-blue shaded area covers
  the whole range of are ratio estimates by \citet{Kreidberg2013}, assuming an average area ratio of 1.344\%. The
  scatter can safely be assumed to be caused by unaccounted-for systematics, and the estimate uncertainties
  are severely underestimated.}
 \label{fig:gj_1215_h_transmission_spectrum}
\end{figure}

The results of the analysis are shown in Fig.~\ref{fig:gj_1215_h_transmission_spectrum}. GJ~1214b is known to have an
extremely flat transmission spectrum most likely dominated by clouds \citep{Kreidberg2013}, and we again conclude that 
the scatter seen in the transit depth is not a real feature of the planet's transmission spectrum, and that our 
uncertainties are significantly underestimated.

\section{HD 209458b}
\label{sec:hd_209458}

\subsection{Overview}
\label{sec:hd_209458:overview}

\begin{figure}
 \centering
 \includegraphics[width=\columnwidth]{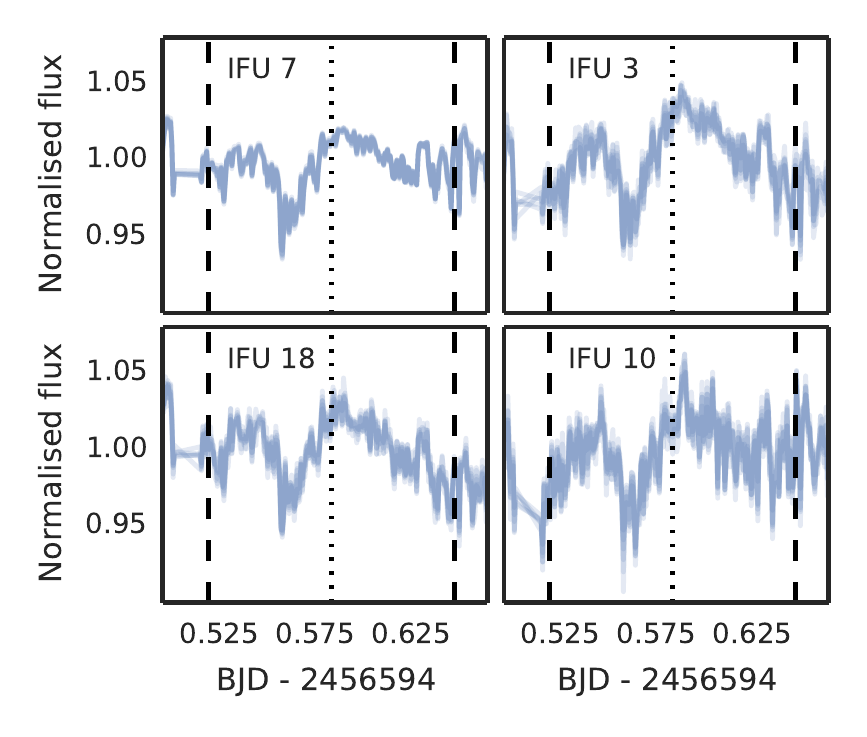}
 \caption{White light curves for HD~209458b (IFU~7) and the three brightest reference stars. The expected transit 
start, center, and end times are marked as vertical dotted lines.}
 \label{fig:hd_209458b_relative_fluxes}
\end{figure}

The HD~209458 observations included pre-imaging to ensure that all the stars were well centered in the IFUs 
(Fig.~\ref{fig:hd_209458b_target_images}). The centering is good, but the number of pre- and post-transit baseline 
exposures is limited due to technical issues and observational limitations. The flux loss due to the seeing 
variations (from 0.8\arcsec{} to 1.4\arcsec{}) is again the main source of systematics, but this can be accounted for 
to a degree using a PSF-modeling based flux-loss model and Gaussian processes. The transit is not visible in the raw 
light curve (Fig.~\ref{fig:hd_209458b_relative_fluxes}), and the quality of the data is lower than what could be 
expected based on the WASP-19 observations.

\begin{figure*}
 \centering
 \includegraphics[width=\textwidth]{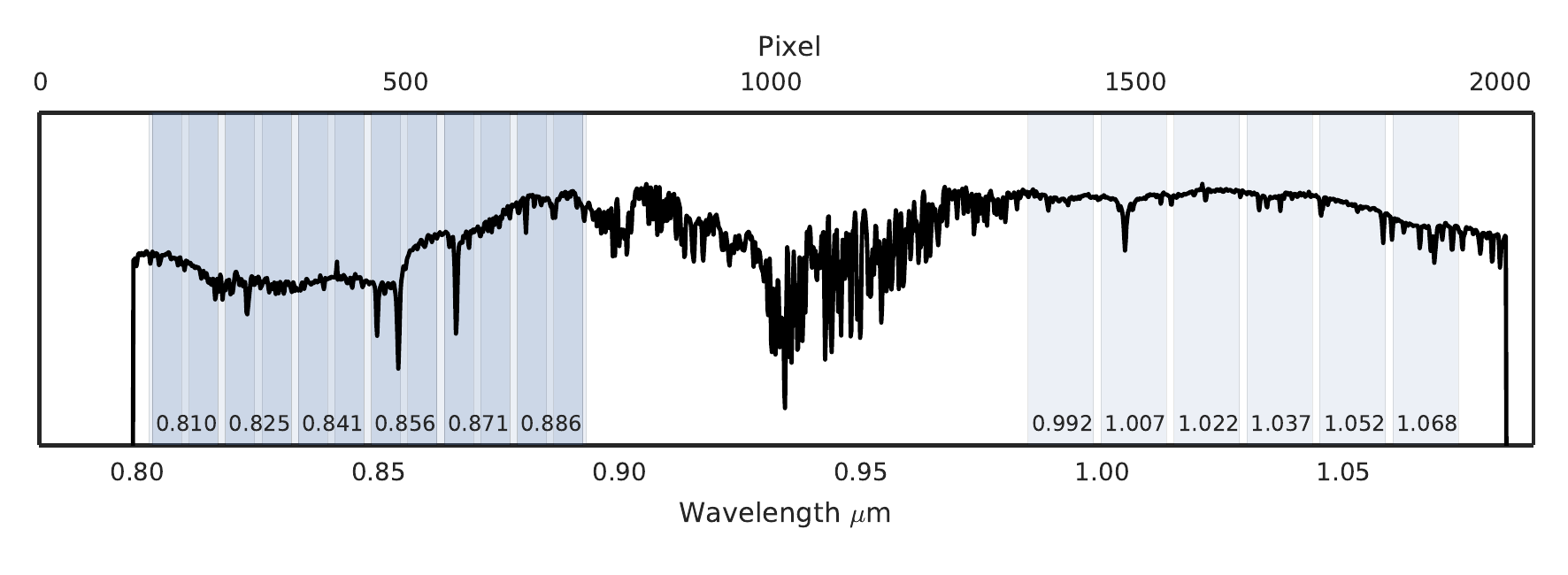}
 \caption{HD~209458b spectrum with the $i$ and $z$ bands marked with light blue, and 6 narrow passbands for each marked as darker blue. 
  A separate analysis was carried out for the $i$ band divided into 12 narrow passbands.}
 \label{fig:hd_209458b_spectrum_bins}
\end{figure*}

\subsection{Analysis and Results}
\label{sec:hd_209458:analysis}

We begin by investigating how well the systematics can be modeled using modeled flux losses and Gaussian processes. 
First, we calculate the flux losses for each star used in the analysis based on PSF fitting. We continue by estimating 
the fractional flux loss not corrected by the relative photometry, shown in Fig.~\ref{fig:hd_209458b_aux_pars}, and use 
this as an input parameter to a Gaussian process with a squared exponential kernel.

After accounting for the flux loss, the light curve contains still clear systematics from varying sky background. We 
include the sky level as a second GP input parameter, and end up with a GP characterized by two additive squared 
exponential kernels, each with an independent input scale parameter. We optimize the three GP hyperparameters (output 
scale, two input scales), and keep them fixed during the MCMC run.

\begin{figure}
 \centering
 \includegraphics[width=\columnwidth]{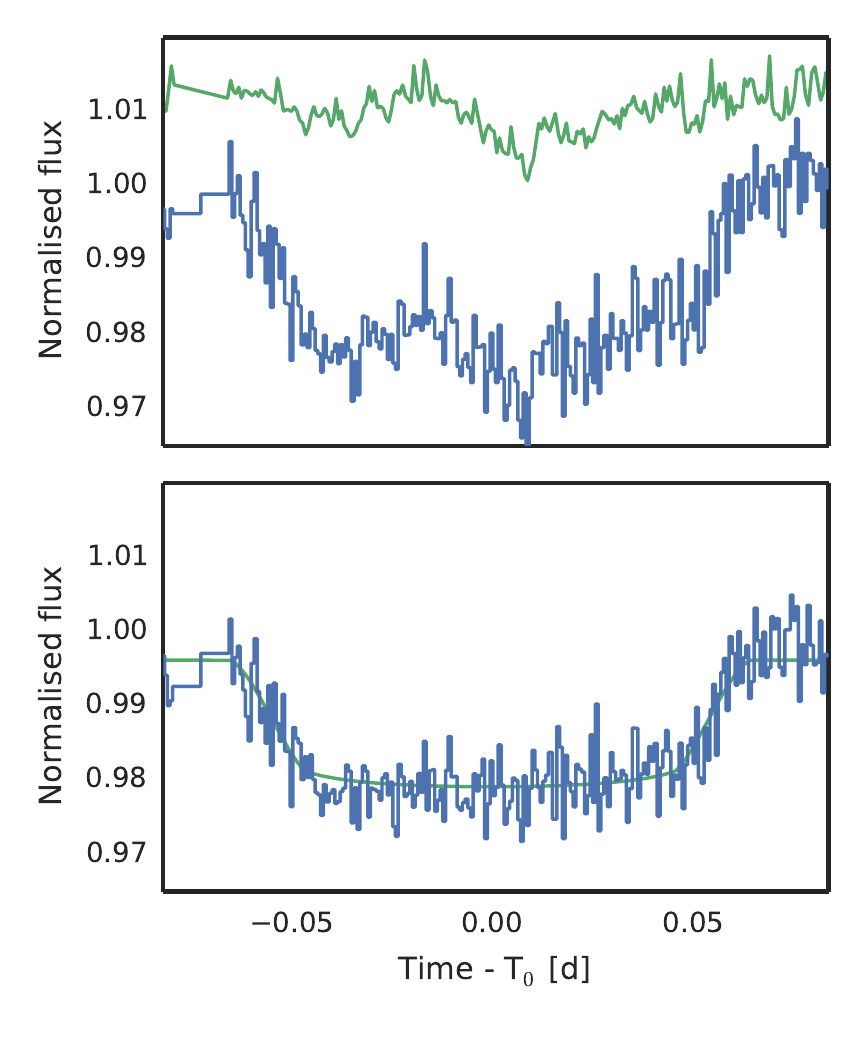}
 \caption{Top figure: HD~209458b z-band light curve divided by the sum of the fluxes from the three brightest reference 
stars (blue line) and the mean systematics explained by a Gaussian process model (green). Bottom figure: The relative 
photometry corrected with the mean systematics (blue) and the fitted transit model (green).}
 \label{fig:hd_209458b_relative_lc}
\end{figure}

\begin{figure*}
 \centering
 \includegraphics[width=\textwidth]{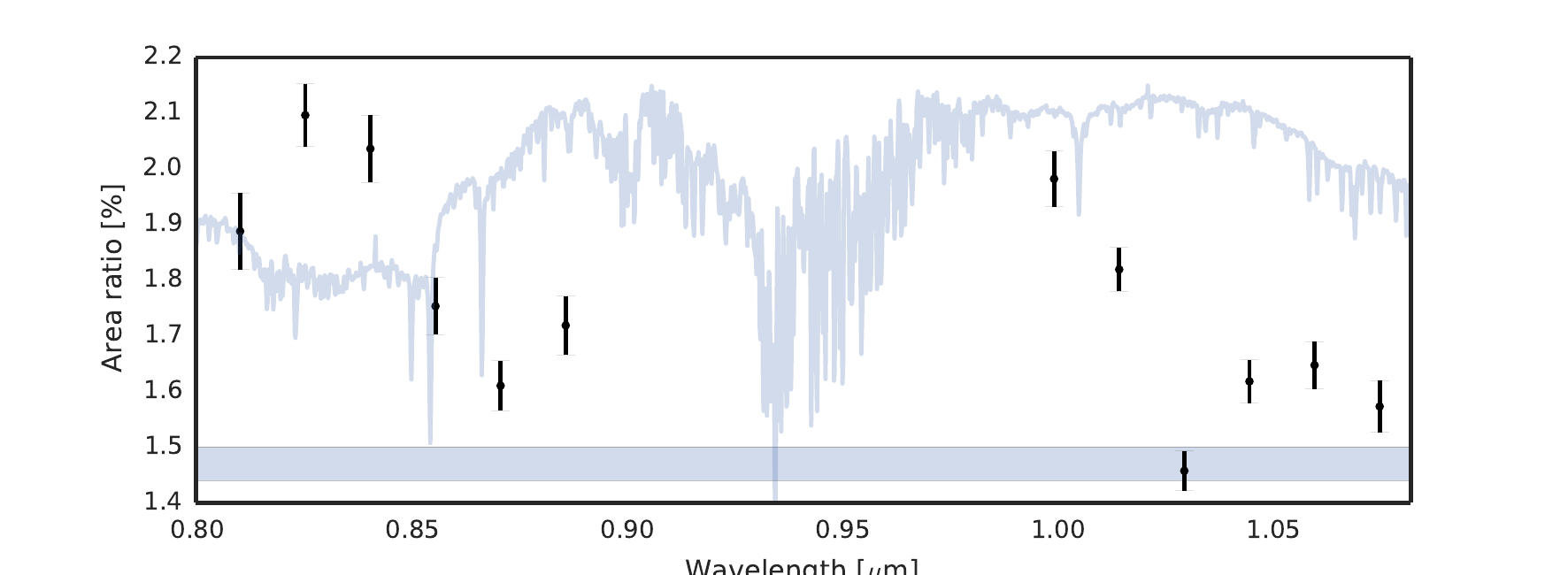}
 \caption{HD~209458b $i$ and $z$ transmission spectrum. The blue band shows the maximum expected variation in the are 
  ratio.}
 \label{fig:hd_209458b_final}
\end{figure*}

The final $z$ broadband light curve, the GP-based systematics, and the systematics-corrected light curve are shown in 
Fig.~\ref{fig:hd_209458b_relative_lc}. Our broadband light curve rms scatter estimate of 3.5$\times 10^{-3}$
is almost 20 times the theoretical shot noise estimate of $\sim2\times10^{-4}$. The GP using the estimated sky level
and the results from the flux loss modeling is capable of explaining most of the systematics. However, the average
radius ratio estimate for the wide $z$-band disagrees with the previous studies, and the transmission spectrum for $i$
and $z$ bands, shown in Fig.~\ref{fig:hd_209458b_final}, shows scatter with a significantly larger amplitude than what
expected from theory \citep[or observed previously, for example][]{Sing2008,Evans2015}. It is clear that our
baseline observations are not sufficient to constrain the transit depth, and our parameter estimate uncertainties are
again underestimated.

\section{Conclusions}

The test based on the observations of the partial transit of WASP-19b gives cautious support that KMOS can, in optimal
conditions, deliver the photometric precision needed for transmission spectroscopy. However, the observations of
GJ~1214b and HD~209458b teach us that the targets must be highly favorable (bright stars with several comparison stars
of similar magnitude located within the patrol field), the IFU centering must be exact (which can be difficult for
targets with high proper motion) favorable targets, the observation conditions need to be stable due to the issues with
flux loss, and sufficient out-of-transit baseline must be obtained. These requirements combined with time-critical
nature of transit observations---and the need to repeat the observations several times---lead to rigid constraints on
telescope scheduling, and the sensitivity on seeing variations means that the risk of not obtaining useful data is
high. 

Finally, the fact that similar or better quality can routinely be achieved with telescopes and instruments with smaller
observation pressure \citep{Bean2013,Bean2010a,Bean2011,Croll2011,Crossfield2011}, and the fact that transmission
spectroscopy does not gain significantly from having spatially resolved spectra, we conclude that while transmission
spectroscopy can be carried out with KMOS, it is not the optimal science case to take advantage of the abilities offered
by the instrument and a 8.2~m telescope. If one nevertheless considers a case where the benefits can be larger than the
risks, one should 
\begin{itemize}
  \item do pre-imaging to ensure that all the stars are well centered in their IFUs,
  \item ensure that all the target IFUs have good sky IFUs on the same detector as the targets,
  \item ensure that a proper OOT baseline is observed.
\end{itemize}

\section*{Acknowledgements}
We warmly thank the anonymous referee for their useful comments and recommendations.
HP has received support from the Leverhulme Research Project grant RPG-2012-661.

\bibliographystyle{mn2e}
\bibliography{kmos}
\end{document}